\documentclass[11pt]{article}

\usepackage{amssymb,latexsym,amsmath}
\usepackage{xcolor}
\usepackage{bm}

\usepackage{hyperref}

\begin{document}

\newtheorem{theorem}{Theorem}[section]

\newtheorem{proposition}[theorem]{Proposition}

\newtheorem{lemma}[theorem]{Lemma}

\newtheorem{corollary}[theorem]{Corollary}

\newtheorem{definition}[theorem]{Definition}

\newtheorem{remark}[theorem]{Remark}

\newtheorem{exempl}{Example}[section]

\newenvironment{exemplu}{\begin{exempl}  \em}{\hfill $\surd$

\end{exempl}}

\newcommand{\ea}{\mbox{{\bf a}}}
\newcommand{\eu}{\mbox{{\bf u}}}
\newcommand{\ep}{\mbox{{\bf p}}}
\newcommand{\ed}{\mbox{{\bf d}}}
\newcommand{\eD}{\mbox{{\bf D}}}
\newcommand{\eK}{\mathbb{K}}
\newcommand{\eL}{\mathbb{L}}
\newcommand{\eB}{\mathbb{B}}
\newcommand{\ueu}{\underline{\eu}}
\newcommand{\ueo}{\overline{u}}
\newcommand{\oeu}{\overline{\eu}}
\newcommand{\ew}{\mbox{{\bf w}}}
\newcommand{\ef}{\mbox{{\bf f}}}
\newcommand{\eF}{\mbox{{\bf F}}}
\newcommand{\eC}{\mbox{{\bf C}}}
\newcommand{\en}{\mbox{{\bf n}}}
\newcommand{\eT}{\mbox{{\bf T}}}
\newcommand{\eV}{\mbox{{\bf V}}}
\newcommand{\eU}{\mbox{{\bf U}}}
\newcommand{\ev}{\mbox{{\bf v}}}
\newcommand{\eve}{\mbox{{\bf e}}}
\newcommand{\uev}{\underline{\ev}}
\newcommand{\eY}{\mbox{{\bf Y}}}
\newcommand{\eP}{\mbox{{\bf P}}}
\newcommand{\eS}{\mbox{{\bf S}}}
\newcommand{\eJ}{\mbox{{\bf J}}}
\newcommand{\leb}{{\cal L}^{n}}
\newcommand{\eI}{{\cal I}}
\newcommand{\eE}{{\cal E}}
\newcommand{\hen}{{\cal H}^{n-1}}
\newcommand{\eBV}{\mbox{{\bf BV}}}
\newcommand{\eA}{\mbox{{\bf A}}}
\newcommand{\eSBV}{\mbox{{\bf SBV}}}
\newcommand{\eBD}{\mbox{{\bf BD}}}
\newcommand{\eSBD}{\mbox{{\bf SBD}}}
\newcommand{\ecs}{\mbox{{\bf X}}}
\newcommand{\eg}{\mbox{{\bf g}}}
\newcommand{\paromega}{\partial \Omega}
\newcommand{\gau}{\Gamma_{u}}
\newcommand{\gaf}{\Gamma_{f}}
\newcommand{\sig}{{\bf \sigma}}
\newcommand{\gac}{\Gamma_{\mbox{{\bf c}}}}
\newcommand{\deu}{\dot{\eu}}
\newcommand{\dueu}{\underline{\deu}}
\newcommand{\dev}{\dot{\ev}}
\newcommand{\duev}{\underline{\dev}}
\newcommand{\weak}{\rightharpoonup}
\newcommand{\weakdown}{\rightharpoondown}
\renewcommand{\contentsname}{ }

\title{Dissipation and the information content of the deviation from hamiltonian dynamics}
\author{\href{http://imar.ro/~mbuliga/}{Marius Buliga} \\ Institute of Mathematics of the Romanian Academy}
\date{Version: 27.04.2023. 
Licence: \href{https://creativecommons.org/licenses/by/4.0/}{CC BY 4.0} }


\maketitle

\begin{abstract}
We explain a dissipative version of hamiltonian mechanics, based on the information content of the deviation from hamiltonian dynamics. From this formulation we deduce minimal dissipation principles, dynamical inclusions, or constrained evolution with hamiltonian drift reformulations.  Among applications we recover a dynamics generalization of Mielke et al quasistatic rate-independent processes. 

This article gives a clear and unitary presentation of the theory of hamiltonian inclusions with convex dissipation or symplectic Brezis-Ekeland-Nayroles principle, presented under various conventions first in \cite{bham}  \href{http://arxiv.org/abs/0810.1419}{arXiv:0810.1419}, then in \cite{MBGDS1} \href{https://arxiv.org/abs/1408.3102}{arXiv:1408.3102} and, for the appearance of bipotentials in relation to the symplectic duality, in \cite{binfo} \href{https://arxiv.org/abs/1902.04598v1}{arXiv:1902.04598v1}.  
\end{abstract}

\section{General notations and hamiltonian dynamics}

$\left.\right.$

In hamiltonian mechanics, a physical system is described by a  pair $z = (q,p)$, where  $q\in Q$ is a state vector and $p \in P$ is a momentum vector. The spaces $Q$ and $P$ are topological, locally convex, real vector spaces, with a duality 
$$ (q,p) \mapsto \langle q, p\rangle \in \mathbb{R}$$
The duality is such that for any linear and continuous $\displaystyle A: Q \rightarrow \mathbb{R}$ there is an unique $p \in P$ such that for all $q \in Q$ we have $\displaystyle A(q) = \langle q, p\rangle$. The same is true for the other side: for any linear and continuous $\displaystyle B: P \rightarrow \mathbb{R}$ there is an unique $q \in Q$ such that for all $p \in P$ we have $\displaystyle B(p) = \langle q, p\rangle$.

On the space $Q \times P$ there is a symplectic form, which can be seen as a duality of $Q \times P$ with itself, defined by 
$$ \omega(z',z") = \langle q",p'\rangle - \langle q', p"\rangle$$

The dynamics of the physical system is described via an energy function $H = H(q,p,t)$, called the hamiltonian of the system. We suppose that $H$ is a differentiable function 
$$ H: Q \times P \times \mathbb{R} \rightarrow \mathbb{R}$$
The partial derivatives of $H$ at a point $z = (q,p) \in Q \times P$  are defined via the duality between $Q$ and $P$, so that the partial derivative of $H$ with respect to $p$ is an element of $Q$, with 
$$\langle \frac{\partial H}{\partial p} (q,p,t), p' \rangle \, = \,  \lim_{\varepsilon \rightarrow 0} \frac{1}{\varepsilon} \left( H \left(q, p + \varepsilon p',t \right) - H(q,p,t) \right)$$ 
for any $\displaystyle p' \in P$, and the partial derivative of $H$ with respect to $q$ is an element of $P$, given by 
$$\langle q', \frac{\partial H}{\partial q} (q,p,t) \rangle \, = \,  \lim_{\varepsilon \rightarrow 0} \frac{1}{\varepsilon} \left( H \left(q + \varepsilon q',p,t \right) - H(q,p,t) \right)$$ 
for any $\displaystyle q' \in Q$. 
The derivative of $H$ with respect to the time $t$ is a real number 
$$\frac{\partial H}{\partial t} (q,p,t)  \, = \,  \lim_{\varepsilon \rightarrow 0} \frac{1}{\varepsilon} \left( H \left(q, p, t + \varepsilon \right) - H(q,p,t) \right)$$

The symplectic gradient of $H$, denoted by $\displaystyle XH(q,p,t) \in Q \times P$,  is 
\begin{equation}
XH(q,p,t) = \left(\frac{\partial H}{\partial p}  (q,p,t), -\frac{\partial H}{\partial q}  (q,p,t) \right)
\label{sympgrad}
\end{equation}

The evolution equation for hamiltonian dynamics is: 
$$ \dot{c}(t) \, = \, XH(c(t),t)$$ 
or, in a more clear form  
\begin{equation} \left\{   
\begin{array}{rcl}
\dot{q} & = & \frac{\partial H}{\partial p}  (q,p,t) \\
\dot{p} & = & -\frac{\partial H}{\partial q}  (q,p,t)
\end{array}
\right.
\label{hamilton}
\end{equation} 
where $\displaystyle \dot{q}$, $\displaystyle \dot{p}$ denote derivatives with respect to time of $q$, resp. $p$. 

Hamiltonian dynamics is conservative, in the following sense. Consider an evolution curve $\displaystyle z(t) = (q(t), p(t))$ and compute 
$$\frac{d}{dt} H(z(t),t) - \frac{\partial H}{\partial t}(z(t),t) \, = \, \langle \frac{\partial H}{\partial p} (q,p,t), \dot{p} \rangle + \langle \dot{q}, \frac{\partial H}{\partial q} (q,p,t) \rangle$$
From (\ref{hamilton}) we obtain: 
$$\langle \frac{\partial H}{\partial p} (q,p,t), \dot{p} \rangle + \langle q', \frac{\partial H}{\partial q} (q,p,t) \rangle \, = \, \langle \frac{\partial H}{\partial p} (q,p,t), -\frac{\partial H}{\partial q}  (q,p,t) \rangle + \langle \frac{\partial H}{\partial p}  (q,p,t), \frac{\partial H}{\partial q} (q,p,t) \rangle \, = \, 0$$
Therefore 
$$\frac{d}{dt} H(z(t),t) - \frac{\partial H}{\partial t}(z(t),t) \, = \, 0$$

\section{Likelihoods}

In the symplectic space $\displaystyle Q \times P$ we introduce the maximal likelihood between two vectors as: 
\begin{equation}
\pi_{max}(z',z") \, = \, e^{\min \left\{0, \omega(z',z") \right\}}
\label{maxlike}
\end{equation} 
This is a number in $(0,1]$, for example when $\displaystyle z'$ and $\displaystyle z"$ are colinear their maximal likelihood is 1, or if $\displaystyle  \omega(z',z") > 0$ then again the maximal likelihood is 1. 

\begin{definition}
A likelihood function is a function $\displaystyle \pi: (Q \times P)^{3} \rightarrow [0,1]$ with the properties: for any $\displaystyle z,z',z" \in Q \times P$
\begin{enumerate}
\item[(a)] if either of the  maxima exist: $\displaystyle \max_{v \in Q \times P} \pi(z, z', v)  \, , \, \max_{w \in Q \times P} \pi(z, w, z")$,  then they are equal to $0$ or $1$
\item[(b)] the functions 
$\displaystyle v \in Q \times P \mapsto - \ln \pi(z, z', v) \, , \, w \in Q \times P \mapsto - \ln \pi(z, w, z")$
are convex and lower semicontinuous (lsc).
\end{enumerate} 
The information content of the likelihood function $\pi$ is $\displaystyle I: (Q \times P)^{3} \rightarrow [0,+\infty]$, 
$$I(z,z',z") = - \ln \pi(z,z',z")$$
with the convention that $\displaystyle - \ln 0 = + \infty$. 

A likelihood $\displaystyle \pi: (Q \times P)^{3} \rightarrow [0,1]$ is tempered if moreover for any $\displaystyle z,z',z" \in Q \times P$
\begin{enumerate}
\item[(c)] $ \displaystyle \pi(z,z',z") \leq \pi_{max}(z',z")$
\end{enumerate}
\label{deflikelihood}
\end{definition}
Clearly, the maximal likelihood is a tempered likelihood function according to definition \ref{deflikelihood}, but we shall see that many other, interesting likelihood exist. In order to describe them we need to introduce some convex analysis notions, the classical ones from Moreau \cite{moreau}, adapted to the symplectic space $Q \times P$. 

Let $\displaystyle d: \left(Q \times P\right)^{2} \rightarrow \mathbb{R}$ be a duality of $\displaystyle Q \times P$ with itself. 

\begin{definition}
For a function $\displaystyle f: Q \times P \rightarrow \mathbb{R} \cup \left\{+\infty\right\}$ and a point $\displaystyle (q,p) \in Q \times P$, the left subgradient at $(q,p)$ of $f$ with respect to the duality $d$ is the set $\displaystyle \partial_{d}^{L} f (q,p)$ of all $\displaystyle (q',p') \in Q \times P$ with the property that for any $\displaystyle (q",p") \in Q \times P$ 
$$f(q,p) + d((q',p'), (q",p")-(q,p)) \, \leq \, f(q",p")$$
The left polar of $f$ with respect to the duality $d$ is the function $\displaystyle f^{*L}_{d}: Q \times P \rightarrow \mathbb{R} \cup \left\{+\infty\right\}$, 
$$f^{*L}_{d}(q",p") \, = \, \sup \left\{ d((q",p"),(q',p')) - f(q',p') \, : \, (q',p') \in Q \times P \right\}$$
Likewise, the right subgradient at $(q,p)$ of $f$ with respect to the duality $d$ is the set $\displaystyle \partial_{d}^{R} f (q,p)$ of all $\displaystyle (q',p') \in Q \times P$ with the property that for any $\displaystyle (q",p") \in Q \times P$ 
$$f(q,p) + d((q",p")-(q,p), (q',p')) \, \leq \, f(q",p")$$
The right polar of $f$ with respect to the duality $d$ is the function $\displaystyle f^{*R}_{d}: Q \times P \rightarrow \mathbb{R} \cup \left\{+\infty\right\}$, 
$$f^{*R}_{d}(q",p") \, = \, \sup \left\{ d((q',p'),(q",p")) - f(q',p') \, : \, (q',p') \in Q \times P \right\}$$
\label{convexdef}
\end{definition}

By unwinding this definition we arrive to the following Fenchel inequalities theorem.
\begin{theorem}
(Fenchel inequalities.) Let $\displaystyle f: Q \times P \rightarrow \mathbb{R} \cup \left\{+\infty\right\}$  be convex and lsc. Then for any $\displaystyle (q',p'), (q",p")  \in Q \times P$
$$f(q',p') + f^{*R}_{d}(q",p") \, \geq \, d((q',p'),(q",p"))$$
The function $\displaystyle f^{*R}_{d}$ is convex lsc. The equality 
$$f(q',p') + f^{*R}_{d}(q",p") \, = \, d((q',p'),(q",p"))$$
is equivalent with 
$$(q",p") \in \partial_{d}^{R} f(q',p')$$

Likewise, for any $\displaystyle (q',p'), (q",p")  \in Q \times P$ 
$$f^{*L}_{d}(q',p') + f(q",p") \, \geq \, d((q',p'),(q",p"))$$
The function $\displaystyle f^{*L}_{d}$ is convex lsc. The equality 
$$f^{*L}_{d}(q',p') + f(q",p") \, = \, d((q',p'),(q",p"))$$
is equivalent with 
$$(q',p') \in \partial_{d}^{L} f(q",p")$$

Finally, $\displaystyle g = f^{*R}_{d}$ implies $\displaystyle f = g^{*L}_{d}$, and  $\displaystyle g = f^{*L}_{d}$ implies $\displaystyle f = g^{*R}_{d}$. 
\label{fenchel}
\end{theorem} 

This gives us a way to construct likelihoods. 
\begin{theorem}
For any function $\displaystyle f: Q \times P \rightarrow \mathbb{R} \cup \left\{+\infty\right\}$ which is convex, lsc, and for any duality $\displaystyle d: \left(Q \times P\right)^{2} \rightarrow \mathbb{R}$ be a duality of $\displaystyle Q \times P$ with itself, the following functions  
$$\pi^{R}_{f}(z,z',z") \, = \, \exp \left( d(z',z") -f(z') - f^{*R}_{d}(z") \right)$$
$$\pi^{L}_{f}(z,z',z") \, = \, \exp \left( d(z',z") -f^{*L}_{d}(z') - f(z") \right)$$
are likelihoods in the sense of definition \ref{deflikelihood}. 
\label{sepdlikes}
\end{theorem}

\paragraph{Proof.} Let $\displaystyle I^{R}_{f}$ be the information content of the likelihood $\displaystyle \pi^{R}_{f}$: 
$$I^{R}_{f}(z,z',z") \, = \, f(z') + f^{*R}_{d}(z") - d(z',z")$$
From the definition we see that $\displaystyle I^{R}_{f}(z,z',z")$ is convex and lsc in each of the arguments $\displaystyle z'$ and $\displaystyle z"$. The Fenchel inequality implies that $\displaystyle I^{R}_{f}(z,z',z") \in [0,+\infty]$, equivalently that $\displaystyle \pi^{R}_{f}(z,z',z") \in [0,1]$. If there exists 
$\displaystyle \max_{w \in Q \times P} \pi^{R}_{f}(z, w, z")$ and it is different from $0$, then there exists a $\displaystyle z' \in Q \times P$ such that 
$$\min_{w \in Q \times P} \left(f(w) + f^{*R}_{d}(z") - d(w,z")\right) \, = \,  f(z') + f^{*R}_{d}(z") - d(z',z") \in \mathbb{R}$$
This implies that for any $\displaystyle w \in Q \times P$ we have 
$$ f(w) - d(w,z") \, \geq \, f(z') - d(z',z")$$ 
therefore $\displaystyle z" \in \partial_{d}^{R} f(z')$,  which by Fenchel equality implies $\displaystyle I^{R}_{f}(z,z',z") = 0$. Therefore the $\displaystyle \max_{w \in Q \times P} \pi^{R}_{f}(z, w, z")$ equals $1$. 

Finally, let's denote  $\displaystyle g = f^{*R}_{d}$. Then $\displaystyle f = g^{*L}_{d}$, so 
$$\pi^{R}_{f}(z,z',z") \, = \, \pi^{L}_{g}(z,z',z")$$
which allows us to end the proof, simply by repeating the same reasoning for $g$.  \quad $\square$

For a convex, lsc function $\displaystyle f: Q \times P \rightarrow \mathbb{R} \cup \left\{+\infty\right\}$,  let us denote by 
$$b_{f}(z',z") \, = \, f(z') + f^{*R}_{d}(z")$$
This is called the separable bipotential associated to $\displaystyle f$. Bipotentials were introduced in \cite{gds2} as a convex analysis notion which is well adapted for applications to non-associated constitutive laws. Bipotentials were used in soil mechanics,  plasticity, damage or friction. In all these applications the dualities considered were among static variables, but the definition and theoretical results about bipotentials do apply for any duality.  For the theory of bipotentials and applications in quasistatic mechanics,  see the review paper \cite{gds3}. In \cite{bham} the symplectic duality was used for the first time and the hamiltonian inclusions with dissipation were introduced and applied to dynamic damage mechanics. We showed there that we obtain a generalization in dynamics of Mielke et al theory of rate-independent processes \cite{mielke}, \cite{mielketh99}, \cite{mielkethl}. Later this was  continued in \cite{MBGDS1} where separable bipotentials with respect to the symplectic duality were used. The corresponding subgradients and polars were called "symplectic" and the hamiltonian inclusions with dissipation were reformulated as the Symplectic Brezis-Ekeland-Nayroles principle (SBEN)  and used to show that in the quasistatic approximation we can recover classical variational principles of Brezis-Ekeland \cite{Brezis Ekeland 1976} and Nayroles \cite{Nayroles 1976}.

In our context, for a space $\displaystyle Q \times P$ which is in duality $\displaystyle d: \left(Q \times P\right)^{2} \rightarrow \mathbb{R}$ with itself, bipotentials have the following definition. 

\begin{definition}
A function $\displaystyle b: \left(Q \times P\right)^{2} \rightarrow \mathbb{R} \cup \left\{+\infty\right\}$ is a bipotential if: 
\begin{enumerate}
\item[(a)] $\displaystyle b(z',z") \geq d(z',z")$ for any $\displaystyle z',z" \in Q \times P$
\item[(b)] $\displaystyle b(z',z") =  d(z',z")$ if and only if $\displaystyle z" \in \partial_{d}^{R} b(\cdot,z")(z')$ if and only if $\displaystyle z' \in \partial_{d}^{L} b(z',\cdot)(z")$
\end{enumerate}
\label{bipodef}
\end{definition}

Likelihoods are related to bipotentials. The relation has been noted before, where information contents of likelihoods appear as syncs, definition 2.3 \cite{gds4}. The same was first observed in relations (51), (52) \cite{laborderenard}. The proof of the following theorem is the same as the one of proposition 2.4 \cite{gds4}, only adapted to the notations of the present article. 

\begin{theorem}
Let $\displaystyle \pi: (Q \times P)^{3} \rightarrow [0,1]$ be a function and $\displaystyle d: \left(Q \times P\right)^{2} \rightarrow \mathbb{R}$ be a duality of $\displaystyle Q \times P$ with itself. Denote by $\displaystyle I = - \ln \pi$ the information content of the function $\displaystyle \pi$, and by 
$$b_{d}(z,z',z") = I(z,z',z") + d(z',z")$$
The function $\displaystyle \pi$ is a likelihood if and only if the function $\displaystyle b_{d}(z,\cdot,\cdot)$ is a bipotential with respect to the duality $\displaystyle d$.  
\label{likebipo}
\end{theorem}

As we can see, while likelihoods are independent of dualities, bipotentials are relative to a duality. We can easily transform a bipotential $\displaystyle b$ with respect to the duality $\displaystyle d$ into another bipotential $\displaystyle b'$ with respect to another duality $d'$, by the formula: 
$$b'(z',z") - d'(z',z") \, = \, b(z',z") - d(z',z")$$

For the particular duality $\displaystyle \omega$, the symplectic form, to any likelihood we associate its symplectic bipotential. 

\begin{definition}
Let $\displaystyle \pi: (Q \times P)^{3} \rightarrow [0,1]$ be a likelihood. 
With $\displaystyle I(z,z',z") = - \ln \pi(z,z',z")$ the information content of $\displaystyle \pi$, the symplectic bipotential associated to this likelihood is
$$b_{\omega}^{\pi}(z,z',z") \, = \, I(z,z',z") + \omega(z',z")$$

The minimal symplectic bipotential is the symplectic bipotential of the maximal likelihood $\displaystyle \pi_{max}$, i.e.
$$b_{\omega}^{min}(z',z") \, = \,  \max \left\{0, \omega(z',z") \right\}$$
\label{bsimp}
\end{definition}

A likelihood $\displaystyle \pi$ is tempered if and only if for any $\displaystyle z, z',z" \in Q \times P$ 
$$b_{\omega}^{\pi}(z,z',z") \geq 0$$
or equivalently 
$$b_{\omega}^{\pi}(z,z',z") \geq b_{\omega}^{min}(z',z")$$

\section{Deviation from hamiltonian dynamics}

We give here a dissipative modification of hamiltonian mechanics (\ref{hamilton}), which continues  \cite{binfo}, \cite{bham}, \cite{MBGDS1}, 
\cite{sben1}. In this modification we need a hamiltonian and a tempered likelihood function.

The dynamics of a disipative physical system is obtained by the introduction of new variables, gathered in the gap vector  $\displaystyle \eta = (\eta_{q}, \eta_{p}) \in Q \times P$. We shall also need new equations, which come from the likelihood function.  

\begin{definition}
Given a hamiltonian $H$ and a tempered likelihood function $\pi$ 
$$\displaystyle H: Q \times P \times \mathbb{R} \rightarrow \mathbb{R} \quad , \quad \pi: (Q \times P)^{3} \rightarrow [0,1]$$ 
the dynamics of a  physical system is defined by the modification of the hamiltonian dynamics equations (\ref{hamilton}) 
\begin{equation} \left\{   
\begin{array}{rcl}
\dot{q} & = & \frac{\partial H}{\partial p}  (q,p,t) \, + \, \eta_{q} \\
\dot{p} & = &  - \frac{\partial H}{\partial q}  (q,p,t) \, + \, \eta_{p} 
\end{array}
\right.
\label{hamiltongap}
\end{equation} 
together with the new equation:  
\begin{equation}
\pi(z, \dot{z}, \eta) \, = \, 1
\label{likely}
\end{equation} 
with the notations $\displaystyle z=(q,p), \eta = (\eta_{q}, \eta_{p}) \in Q \times P$, $\displaystyle \dot{z} = (\dot{p}, \dot{q}) \in Q \times P$. 
\label{mainproblem}
\end{definition}

Let's put the equation (\ref{likely}) in a more explicit form. For a duality $\displaystyle d$, we know from theorem \ref{likebipo} that 
$$b_{d}(z,z',z") = I(z,z',z") + d(z',z")$$
is a bipotential with respect to $\displaystyle d$, where $\displaystyle I$ is the information content of the likelihood $\displaystyle \pi$. The equation (\ref{likely}) is then equivalent with 
\begin{equation}
\eta \in \partial_{d}^{R} b_{d}(c(t),\cdot,\eta)(\dot{c})
\label{consbipoR}
\end{equation}
which is also equivalent with 
\begin{equation}
\dot{c} \in \partial_{d}^{L} b_{d}(c(t),\dot{c},\cdot)(\eta)
\label{consbipoL}
\end{equation}
We define for any $\displaystyle z \in Q \times P$ and  $\displaystyle t \in [0,T]$ the set $\displaystyle Gap(z,t)$ of all 
$\displaystyle z" \in  Q \times P$ such that 
\begin{equation}
z" \in \partial_{d}^{R} b_{d}(z,\cdot,z")(z"+XH(z,t))
\label{gapeta}
\end{equation}
Remark that $\displaystyle z" \in Gap(z,t)$ if and only if 
$$b_{d}(z,z"+XH(z,t), z") = d(z"+XH(z,t), z")$$
We thus get the following equivalent formulations of the problem \ref{mainproblem}. 
\begin{theorem}
$\displaystyle t \in [0,T] \mapsto (c(t),\eta(t))$ is a solution of \ref{mainproblem} with the initial condition $\displaystyle c(0) = z_{0}$  if and only if the curve $\displaystyle t \in [0,T] \mapsto c(t)$ satisfies $\displaystyle c(0) = z_{0}$ and  any of the following are true for any $\displaystyle t \in [0,T]$  : 
\begin{enumerate}
\item[(a)] (dynamical inclusion) 
\begin{equation}
\dot{c}(t) \in  XH(c(t),t) + \partial_{d}^{R} b_{d}(c(t),\cdot,\dot{c}(t)- XH(c(t),t))(\dot{c}(t))
\label{likelycR}
\end{equation}
\item[(b)] (dynamical inclusion) 
\begin{equation}
\dot{c}(t) \in  \partial_{d}^{L} b_{d}(c(t),\dot{c}(t),\cdot)(\dot{c}(t) - XH(c(t),t))
\label{likelycL}
\end{equation}
\item[(c)] (constraint with hamiltonian drift) 
\begin{equation}
\dot{c}(t) \in XH(c(t),t) + Gap(c(t),t)
\label{likelyeta}
\end{equation}
\item[(d)] (implicit evolution) 
\begin{equation}
b_{d}(c(t),\dot{c}(t),\dot{c}(t) - XH(c(t),t)) \, = \, d(\dot{c}(t),\dot{c}(t) - XH(c(t),t))
\label{bipo0}
\end{equation}
\end{enumerate}
and $\displaystyle \eta$ is defined as: 
\begin{equation}
\eta(t) \, = \, \dot{c}(t) - XH(c(t),t)
\label{hamgapeta}
\end{equation}
\label{formclear}
\end{theorem}

Recall that for the same information content, for different dualities we obtain different bipotentials, therefore 
$$b_{d}(z,z',z") - d(z',z") \, = \, b_{\omega}^{\pi}(z,z',z") - \omega(z',z") \, = \, I(z,z',z")$$

\begin{definition}
The dissipation along a curve $\displaystyle t \in [0,T] \mapsto c(t) \in Q \times P$ is the functional: 
\begin{equation}
Diss^{\pi}(c,0,T) \, = \, \int_{0}^{T} b_{\omega}^{\pi}(\dot{c}(t), \dot{c}(t) - XH(c(t),t))\mbox{ d}t
\label{defdissip}
\end{equation}  
Remark that for any curve 
$$Diss^{\pi}(c,0,T) \, \geq \, 0$$
because the likelihood $\displaystyle \pi$ is tempered. 
\end{definition}

In the following theorem we give the energy balance and  dissipation inequalities. 

\begin{theorem}
Let $\displaystyle t \in [0,T] \mapsto (c(t),\eta(t))$ be a solution of \ref{mainproblem} with the initial condition $\displaystyle c(0) = z_{0}$. Then for any $t \in [0,T]$: 
\begin{enumerate}
\item[(a)](energy balance) 
\begin{equation}
H(c(t),t) \, = \, H(z_{0},0) +  \int_{0}^{t} \frac{\partial H}{\partial t}(c(\tau),\tau)\mbox{ d}\tau \, - \, Diss^{\pi}(c,0,t)
\end{equation}
\item[(b)] (dissipation inequalities)
$$\frac{d}{dt} H(c(t),t)) -  \frac{\partial H}{\partial t}(c(t),t) \, \leq \, 0$$
For any curve $\displaystyle t \in [0,T] \mapsto c'(t)$ which satisfies $\displaystyle c'(0) = c(0)$ we have 
\begin{equation}
Diss^{\pi}(c',0,t) + H(c'(t),t) \, \geq Diss^{\pi}(c,0,t) + H(c(t),t)
\label{mindiss}
\end{equation}
\end{enumerate}
\label{theorembalance}
\end{theorem}

\paragraph{Proof.} From theorem \ref{formclear} (d), the curve $\displaystyle t \in [0,T] \mapsto (c(t),\eta(t))$ be a solution of \ref{mainproblem} with the initial condition $\displaystyle c(0) = z_{0}$ if and only if $\displaystyle \eta$ is given by (\ref{hamgapeta}) and $\displaystyle c$ satisfies 
$$b_{\omega}^{\pi}(\dot{c}(t), \dot{c}(t) - XH(c(t),t)) - \omega(\dot{c}(t), \dot{c}(t) - XH(c(t),t)) \, = \, 0$$
But the same calculation as in the section about hamiltonian dynamics gives 
$$- \omega(\dot{c}(t), \dot{c}(t) - XH(c(t),t)) \, = \, - \omega(XH(c(t),t), \dot{c}(t)) \, = \, \frac{d}{dt} H(c(t),t)) -  \frac{\partial H}{\partial t}(c(t),t)$$
Therefore we obtain 
$$\frac{d}{dt} H(c(t),t)) \, = \, \frac{\partial H}{\partial t}(c(t),t) - b_{\omega}^{\pi}(\dot{c}(t), \dot{c}(t) - XH(c(t),t))$$
We integrate this equality over $[0,t]$ and we obtain the energy balance (a). 

The first dissipation inequality from (b) is a consequence of the positivity of the symplectic bipotential. In order to obtain the second inequality from (b), we introduce the information content gap functional $\displaystyle G(c,0,t)$  for any curve 
 $\displaystyle t \in [0,T] \mapsto c(t)$:
$$G(c,0,t) \, = \, \int_{0}^{t} I(c(\tau), \dot{c}(\tau), \dot{c}(\tau) - XH(c(\tau),\tau))) \mbox{ d}\tau$$
In more detail: 
\begin{equation}
G(c,0,t) =  \int_{0}^{t} I\left( c(\tau),\dot{c}(\tau),\dot{q}(\tau) - \frac{\partial H}{\partial p} (c(\tau),\tau), - \dot{p}(\tau) - \frac{\partial H}{\partial q}  (c(\tau),\tau) \right) \mbox{d}\tau
\label{icgap}
\end{equation}
We compute, for an arbitrary curve: 
$$G(c,0,t) \, = \, \int_{0}^{t} \left[b_{\omega}^{\pi}(c(\tau), \dot{c}(\tau), \dot{c}(\tau) - XH(c(\tau),\tau))) - \omega(\dot{c}(\tau), \dot{c}(\tau) - XH(c(\tau),\tau)) \right]\mbox{ d}\tau$$
$$G(c,0,t) \, = \, Diss^{\pi}(c,0,t) + H(c(t),t) - H(c(0),0)$$
For  a solution $\displaystyle t \in [0,T] \mapsto c(t)$ of \ref{mainproblem} and for any curve $\displaystyle t \in [0,T] \mapsto c'(t)$ which satisfies $\displaystyle c'(0) = c(0)$ , we have: 
$$G(c',0,t) \, \geq \, 0 \, = \, G(c,0,t)$$ 
therefore we obtain a principle of minimal information content disclosed by the curve $\displaystyle c$: 
\begin{equation}
G(c',0,t) \, \geq \, G(c,0,t)
\label{minimalinfo}
\end{equation}
Previous computations show that (\ref{minimalinfo}) is equivalent with 
$$Diss^{\pi}(c',0,t) + H(c'(t),t) - H(c'(0),0) \, \geq \, Diss^{\pi}(c,0,t) + H(c(t),t) - H(c(0),0)$$ 
But $\displaystyle c'(0) = c(0)$, therefore $\displaystyle H(c'(0),0) = H(c(0),0)$. The previous inequality becomes (\ref{mindiss}). \quad $\square$

The inequality (\ref{mindiss}) can be seen as a principle of minimal dissipation. Alternatively, as in \cite{binfo}, we can see this as a principle of minimal information content (\ref{minimalinfo}) disclosed by the deviation from hamiltonian evolution, measured by the  information content gap functional (\ref{icgap}).

\section{Applications}
\label{secex}

\paragraph{Pure dissipative evolution.} We pick the minimal symplectic bipotential:
$$b_{\omega}^{\pi}(z,z',z") \, = \, b_{\omega}^{min}(z',z") \, = \, \max \left\{0, \omega(z',z") \right\}$$
see definition \ref{bsimp}. Then a solution of \ref{mainproblem} in the sense of theorem \ref{formclear} (d) is a $\displaystyle t \in [0,T] \mapsto c(t)$ which satisfies the initial condition $\displaystyle c(0) = z_{0}$ and 
$$b_{\omega}^{min}(\dot{c}(t),\dot{c}(t) - XH(c(t),t)) \, = \, \omega(\dot{c}(t),\dot{c}(t) - XH(c(t),t))$$
This is just
$$\max \left\{0, \omega(\dot{c}(t),\dot{c}(t) - XH(c(t),t)) \right\} \, = \, \omega(\dot{c}(t),\dot{c}(t) - XH(c(t),t))$$
which is equivalent with 
$$\omega(\dot{c}(t),\dot{c}(t) - XH(c(t),t)) \, \geq \, 0$$
Recall that 
$$\omega(\dot{c}(t),\dot{c}(t) - XH(c(t),t)) \, = \, - \frac{d}{dt} H(c(t),t)) +  \frac{\partial H}{\partial t}(c(t),t)$$
therefore any curve $\displaystyle c$ which satisfies the initial condition and has the property
$$\frac{d}{dt} H(c(t),t)) -  \frac{\partial H}{\partial t}(c(t),t) \, \leq \, 0$$
for any $\displaystyle t \in [0,T]$ is a solution of \ref{mainproblem}. 

Let's compute the gap set: $\displaystyle z" \in Gap(z,t)$ if and only if 
$$b_{\omega}^{\pi}(z"+XH(z,t), z") = \omega(z"+XH(z,t), z")$$
which is just 
$$\omega(XH(z,t), z") \, \geq \, 0$$
The theorem \ref{formclear} (c) formulation, i.e. constraint with hamiltonian drift, becomes  
$$ \dot{c}(t) \in XH(c(t),t) + \eta(c(t)) \, , \, \, \omega(XH(c(t),t), \eta(c(t))) \, \geq \, 0$$

\paragraph{Pure Hamiltonian evolution.} Let's pick the information content (\ref{deflikelihood}) to be: 
$$I(z, z', z") \, = \, \chi_{0} (z") = \left\{ \begin{array}{ll}
0 & \mbox{ if } z" = 0 \\ 
+\infty & \mbox{ otherwise } 
\end{array} \right.
$$
This corresponds to a likelihood function: 
$$ \pi(z,z',z") \, = \, \left\{ \begin{array}{ll}
1 & \mbox{ if } z" = 0 \\ 
0 & \mbox{ otherwise } 
\end{array} \right.
$$
The maximization of the likelihood (\ref{likely}) implies that the gap vector $\displaystyle \eta = 0$, therefore the evolution equations (\ref{hamiltongap}) reduce to the pure Hamiltonian evolution equations (\ref{hamilton}).

\paragraph{Dominance.} In general, we may compare the sets of solutions for two symplectic bipotentials 
$$b^{1}_{\omega}(z,z',z") \, \geq \, b^{2}_{\omega}(z,z',z")$$
This corresponds to two likelihoods 
$$\pi^{1}(z,z',z")  \, \leq \, \pi^{2}(z,z',z")$$
It is then easy to see that any solution of \ref{mainproblem} for the likelihood $\displaystyle \pi^{1}$ is also a solution for the same problem, but for the likelihood $\displaystyle \pi^{2}$. We say that $\displaystyle \pi^{2}$ dominates $\displaystyle \pi^{1}$. 

Any tempered likelihood $\displaystyle \pi$ is by definition \ref{deflikelihood} (c) dominated by $\displaystyle \pi_{max}$, so it is not surprizing that any solution of \ref{mainproblem} for the likelihood $\displaystyle \pi$ is also a solution of the same problem, but for the likelihood $\displaystyle \pi_{max}$, i.e. as shown in the previous example, it dissipates. 

\paragraph{Smooth dissipation.} We consider a symplectic bipotential as in theorem \ref{sepdlikes}
$$b_{\omega}(z,z',z") \, = \, f(z') +  f^{*R}_{\omega}(z")$$
where $\displaystyle f$ is a smooth positive convex function. Mind that we have to choose $\displaystyle f$ such that $\displaystyle b_{\omega}$ is nonnegative! Suppose we are in this situation, let's see what are the equations of \ref{mainproblem} in this case. We use theorem \ref{formclear}. The dynamical inclusion (a) becomes 
$$\dot{c}(t) \in  XH(c(t),t) + \partial_{\omega}^{R}f(\dot{c}(t))$$
But $\displaystyle z" \in \partial_{\omega}^{R}f(z')$ if and only if for any $\displaystyle z \in Q \times P$ we have 
$$f(z') + \omega(z-z',z") \, \leq \, f(z)$$
which is equivalent with 
$$f(z') - \omega(z',z") \, \leq \, f(z) - \omega(z,z")$$
Because $\displaystyle f$ is smooth, this is equivalent with 
$$z" \, = \, Xf(z')$$
Therefore our equation becomes 
\begin{equation}
\dot{c}(t) \, = \,  XH(c(t),t) + Xf(\dot{c}(t))
\label{smoothdis}
\end{equation}
For simplicity let us suppose that $\displaystyle f(z) \, = \, \phi(q)$. Then 
$$Xf(\dot{c}(t)) \, = \, (0, - \frac{\partial \phi}{\partial q}(\dot{q}(t)))$$
Let us compute also $\displaystyle f^{*R}_{\omega}(z")$ in this case, where $\displaystyle z" = (q",p")$: 
$$f^{*R}_{\omega}(q",p") \, = \, \sup \left\{ \langle q",p \rangle + \langle q, -p"\rangle - \phi(q) \, : \, z = (q,p) \in Q \times P \right\}$$
$$f^{*R}_{\omega}(q",p") \, = \, \chi_{0}(q") + \phi^{*}(-p")$$
where $\displaystyle \phi^{*}$ is the usual polar with respect to the duality between $Q$ and $P$. The symplectic bipotential is then 
$$b_{\omega}((q',p'),(q",p")) \, = \, \phi(q') + \phi^{*}(-p") + \chi_{0}(q")$$
and $\displaystyle b_{\omega}((q',p'),(q",p")) \, \geq \, 0$ for all $\displaystyle z', z" \in Q \times P$  is equivalent with 
$$\phi(q) + \phi^{*}(p) \, \geq \, 0$$
for all $\displaystyle q \in Q, \, p \in P$. In the familiar case where $Q$ and $P$ are the same Hilbert space  with norm $\displaystyle \| \cdot \|$, if we pick $\displaystyle \phi(q) \, = \, \frac{a}{2} \|q \|^{2}$ (for some $a>0$) then $\displaystyle \phi^{*}(-p)  \, = \, \frac{1}{2a} \|p \|^{2}$ and the symplectic bipotential is nonnegative. We discover Rayleigh dissipation: 
$$ \left\{   
\begin{array}{rcl}
\dot{q} & = & \frac{\partial H}{\partial p}  (q,p,t)  \\
\dot{p} & = &  - \frac{\partial H}{\partial q}  (q,p,t) \, - \, \frac{\partial \phi}{\partial q}(\dot{q}) 
\end{array}
\right.
$$

\paragraph{On discontinuous solutions.} For nonsmooth symplectic bipotentials we might need to reformulate \ref{mainproblem} in order to accept discontinuous solutions. Indeed, for example in contact problems (also in plasticity or damage, as shown in \cite{binfo}, \cite{bham}) the solution curve $\displaystyle c(t) = (q(t),p(t))$ may be continuous in the position variable $q$ but discontinuous in the momentum variable $p$, or alternatively some of the variables are continuous and some are discontinuous. In order to cover such cases we have to pick a weak formulation of (\ref{hamiltongap}), in the sense of distributions. A good choice seems to be to pick $\displaystyle c$ as a function of bounded variation over $[0,T]$ and $\eta$ a measure with singular part with respect to $\mbox{ d}t$ concentrated on the jump set of $\displaystyle c$. We leave these technicalities for another paper, even if they are significant in the case of the next application. 

\paragraph{Relation with rate-independent processes.} We show that we can cover a dynamical version of Mielke and collaborators -- Mielke, Theil 
\cite{mielketh99},   Mielke, Theil and Levitas \cite{mielkethl},  \cite{mielke} --  quasistatic rate-independent evolutionary processes. This was done first time in \cite{bham}, but here we can give a short and useful description.  

We pick a symplectic bipotential of the form: 
$$b_{\omega}(z,z',z") \, = \, f(z') +  f^{*R}_{d}(z")$$
as in the smooth dissipation example, but now $\displaystyle f$ is convex, positive and 1-homogeneous: for any positive scalar  $\displaystyle a>0$ 
$$f(a z') \, = \, a f(z')$$
The function $f$ is no longer smooth (because not derivable in $0$, at least). For the dual  
$$f^{*R}_{\omega}(z") \, = \, \sup \left\{ \omega(z,z")- f(z) \, : \, z \in Q \times P \right\}$$
notice that for any $\displaystyle a>0$ 
$$f^{*R}_{\omega}(z") \, = \, \sup \left\{ \omega(az,z")- f(az) \, : \, az \in Q \times P \right\} \, = \, a f^{*R}_{d}(z")$$
therefore $\displaystyle f^{*R}_{\omega}(z") \in \left\{ 0, + \infty \right\}$, i.e. it is a characteristic function of a set $\displaystyle C \in Q \times P$ with $\displaystyle 0 \in C$, more precisely: 
\begin{equation}
C \, = \, \left\{ z" \in Q \times P \mbox{ : } \omega(z,z") \, \leq \, f(z) \, \forall z \in Q \times P \right\}
\label{supportset}
\end{equation}
 Therefore 
$$b_{\omega}(z,z',z") \, = \, f(z') +  \chi_{C}(z")$$
and the equations for the problem \ref{mainproblem} are, in terms of gap sets: 
\begin{equation}
Gap(z,t) \, = \, \left\{ z" \in C \mbox{ : } f(z" + XH(z,t)) = \omega(XH(z,t),z") \right\}
\label{gaphom}
\end{equation}
\begin{equation}
\dot{c}(t) - XH(c(t),t) \in Gap(c(t),t)
\label{dynhom}
\end{equation}
$\displaystyle f$ is a dissipation potential and the dissipation functional (\ref{defdissip}) is 
$$Diss^{\pi}(c,0,T) \, = \, \int_{0}^{T} f(\dot{c}(t)) + \chi_{C}(\dot{c}(t) - XH(c(t),t))\mbox{ d}t$$
Theorem \ref{theorembalance} applied for this case gives us the energy balance equations, dissipation inequalities and the principle of minimal dissipation, thus it allows us to extend the formulation of Mielke et al rate-independent processes, but this time in dynamics.

\end{document}